\documentclass[aps,twocolumn]{revtex4}
\usepackage{bm}
\usepackage{amsmath}
\usepackage{amssymb}
\usepackage{graphicx}
\usepackage{epsfig}
\usepackage{epstopdf}
\usepackage{mathrsfs}

\begin{document}
\newtheorem{theorem}{Theorem}
\newtheorem{corollary}{Corollary}

\def\be{\begin{equation}}
\def\en#1{\label{#1}\end{equation}}

\newcommand{\rd}{\mathrm{d}}
\newcommand{\vare}{\varepsilon }
  \newcommand{\tb}{\mathbf{t}}
\newcommand{\Phib}{\mathbf{\Phi}}
\newcommand{\Psib}{\mathbf{\Psi}}
  \newcommand{\U}{\mathcal{U}}
 \newcommand{\cH}{\mathcal{H}}

\title{Exact State Evolution and Energy Spectrum in Solvable Bosonic Models}

 \author{Valery  Shchesnovich} 
 \email{ valery@ufabc.edu.br }
 
\affiliation{Centro de Ci\^encias Naturais e Humanas, Universidade Federal do
ABC, Santo Andr\'e,  SP, 09210-170 Brazil }

\begin{abstract}
Solvable bosonic models provide a fundamental framework for describing light propagation in nonlinear media, including optical down-conversion processes that generate squeezed states of light and their higher-order generalizations. In   quantum optics a central objective is to determine the time evolution of a given initial state.   Exact analytic solution to the state-evolution problem is presented, applicable to  a broad class of solvable bosonic models and arbitrary initial states.  Moreover, the characteristic equation governing the energy spectrum is  derived and the   eigenstates are found  in the form of  continued fractions  and as the principal  minors of the associated Jacobi matrix. The results provide a solid analytical framework  for discussion of exactly solvable bosonic models. \end{abstract}
\maketitle

\section{Introduction}
For a broad class of bosonic models describing interactions among several bosonic modes, the Hilbert space decomposes into a direct sum of finite-dimensional invariant subspaces. Within each subspace, the Hamiltonian assumes the form of a tridiagonal Hermitian matrix. These two structural features — the decomposition of the Hilbert space into finite-dimensional invariant subspaces and the tridiagonal representation of the Hamiltonian therein — characterize a class of exactly solvable bosonic models whose energy spectra has been  studied  by  the  group-theoretic techniques \cite{GroupMeth,LieAlg3Boson,GroupMeth2,LieAlgGen1,LieAlgGen2}, by employing  the   Bethe ansatz \cite{SpinSystems,3BosonBethe}, or by the Quantum Inverse Scattering Method \cite{QISM}. 

In quantum optics, the propagation of optical modes in a lossless nonlinear medium, under the condition of phase matching, can be described by a solvable bosonic model. In a lossless medium, energy conservation is governed by the Manley–Rowe relations \cite{NonlOpt}, which ensure that the total optical energy remains constant during propagation. Accordingly, the free-propagation term $\hat{H}_0$ of the quantum Hamiltonian, representing the optical energy, is quadratic in the bosonic creation and annihilation operators, while the interaction term $\hat{H}$, of higher order in these operators, governs the mode-conversion processes mediated by the nonlinear medium. When the phase-matching condition is satisfied, the interaction term preserves the total optical energy and therefore commutes with $\hat{H}_0$,
$[\hat{H}_0, \hat{H}] = 0 $.
Prominent examples include $k$-photon down-conversion processes \cite{GenSqueez,3phDC}, which generalize the second-order down-conversion mechanism responsible for the generation of twin photons and squeezed states of light \cite{ReviewSq,Couteau18,Zhang21}.

In quantum-optical applications of solvable bosonic models, the central problem is the determination of the time evolution of the system’s quantum state. Under strong pump condition, the so-called parametric approximation is commonly employed, in which the bosonic operators corresponding to the pump mode are replaced by complex scalars. This approximation leads to the well-known description of squeezed states of light \cite{ReviewSq}. Beyond the parametric approximation, however, the analysis of solvable bosonic models generally requires more advanced algebraic frameworks, such as deformed Lie algebras or the Quantum Inverse Scattering Method, both of which involve substantial mathematical complexity. Consequently, most physical treatments rely on approximate approaches, including WKB-type techniques \cite{WKB4GenFun,Hillery84,WKBpump,Crouch88} and numerical simulations \cite{Reid88,Drobny92,Buzek93,Drobny94,Hillery95}, often supported by reductions based on conservation laws. Extending the analysis beyond the parametric approximation is of particular importance, since this regime is expected to exhibit pronounced non-Gaussian quantum effects \cite{BeyondPA1,BeyondPA2}.

Recently, a simple algebraic method was found \cite{My1} for obtaining the exact solution of the state-evolution problem in a broad class of solvable bosonic models.  The method is particularly applicable to systems arising in quantum-optical contexts, such as $k$-photon down-conversion and related processes, for which the explicit time evolution was derived for a specific initial state in Ref.~\cite{My1}. The purpose of the present work is to apply that approach  to find  the evolution of an arbitrary initial state and to derive explicit expressions for the corresponding eigenstates and for the characteristic equation determining the energy spectrum.

The structure of the paper is as follows. In Section~\ref{sec2}, the class of bosonic models to which the present approach applies is outlined. Section~\ref{sec3} provides the solution of the state-evolution problem for an arbitrary initial state by giving  the average of the unitary evolution operator between two arbitrary basis states within each invariant subspace of the Hilbert space. Theorem1 and Corollaries1 and 2 summarize the main results of this section. In Section~\ref{sec4}, the characteristic polynomial determining the energy eigenvalues in each invariant subspace is derived, and the quantum amplitudes of the corresponding eigenstates are expressed in terms of continued fractions and as  Jacobi determinants. Theorem 2 and Corollary 3 contain the main results concerning the energy spectrum. Furthermore, the stationary state (the eigenstate with zero energy) is obtained in explicit analytic form. Section \ref{sec5} discusses the importance of the results and their possible use in applications. Finally, Section~\ref{sec6} presents a brief summary of the results and points on an open problem.
\section{Solvable   bosonic   models   }
\label{sec2}

 The solvable bosonic models considered here are characterized by two main features. In the interaction picture, they can be described as follows (for further details, see Ref.~\cite{My1}):

\noindent\textit{(i) Invariant partition of the Hilbert space.}—The (infinite-dimensional) Hilbert space $\mathcal{H}$ decomposes into a direct sum of finite-dimensional subspaces, each invariant under the action of the interaction Hamiltonian $\hat{H}$. One can  label the invariant subspaces by their  dimension $N+1$:
 \be
 \cH = \bigoplus_{N} \mathcal{H}_N,\quad \cH_N = \mathrm{Span}\{|\Psi^{(N)}_0\rangle,|\Psi^{(N)}_1\rangle,\ldots,|\Psi^{(N)}_{N}\rangle  \}.
 \en{H_N}
 Without loss of generality, if multiple subspaces share the same dimension, an additional label, e.g., $\mathcal{I} = (N,\ell)$, can be introduced to distinguish them. In the following, we focus on a single invariant subspace $\mathcal{H}_N$, and therefore omit the subspace indices.

 \noindent\textit{(ii) Ladder-type structure of the interaction Hamiltonian.}—The interaction Hamiltonian can be expressed as the sum of two Hermitian-conjugate operators,
\be
\hat{H}=\hat{A}+\hat{A}^\dag,
\en{H}  
which generate transitions between nearest-neighbor basis states within each invariant subspace:
\be
 \langle\Psi_n| \hat{A}|\Psi_{m}\rangle = \langle\Psi_{m}| \hat{A}^\dag|\Psi_{n}\rangle=0, \quad m\ne n+1. 
 \en{GTP2}
It is always possible to choose the phases of the basis states in Eq.~(\ref{H_N}) such that all nonzero matrix elements are non-negative scalars (generally distinct for each invariant subspace). With this choice, the ladder operators act as
 \be 
\hat{A}|\Psi_{n+1}\rangle = \sqrt{\beta_n} |\Psi_{n}\rangle, \quad \hat{A}^\dag|\Psi_n\rangle= \sqrt{\beta_n}|\Psi_{n+1}\rangle,
\en{Psi_n}
where     $\beta_n\ge 0$  and $\beta_N = 0$   in subspace $\cH_N$  due to the finite dimension $\mathrm{dim}\cH_N=N+1$.

As discussed in Ref.\cite{My1}, many bosonic models describing interactions among two or more optical modes in lossless nonlinear media — where the total optical energy of the interacting modes is conserved — possess the two features outlined above. Exactly integrable bosonic models, whose energy spectra have been analyzed in Refs.~\cite{GroupMeth,GroupMeth2,LieAlg3Boson,LieAlgGen1,LieAlgGen2}, also belong to this class.
 
For a  concrete example, consider a sub-class of nonlinear bosonic models characterized by the ladder operator
\be 
  \hat{A}   =  (\hat{a}^\dag)^m \prod_{s=1}^S \hat{b}_s^{k_s},
 \en{models}
where $m\ge 1$, $S\ge 1$  and $k_s\ge 1$  are arbitrary integers. In  each invariant  subspace of Eq. (\ref{H_N})  the state $|\Psi_0\rangle$  is  annihilated by the ladder operator: $\hat{A}|\Psi_0\rangle = 0$.  In the simplest two-mode case ($S=1$) with  $\hat{A} = (\hat{a}^\dag)^m\hat{b}^k$, the invariant subspaces are labeled by the composite index  $\mathcal{I} \equiv  (N,\ell)$, where   $0\le \ell \le k-1$. In each subspace  $\cH_\mathcal{I}$, the basis states $|\Psi^{(\mathcal{I})}_n\rangle$ are given by the Fock states of the two modes:
\begin{eqnarray}
\label{model_1}
&& |\Psi^{(\mathcal{I})}_n\rangle \equiv |M-mn,kn+\ell \rangle, \quad  0\le n\le N\equiv \left[ \frac{M}{m}\right] ,\nonumber\\
&&  |M-mn,kn+\ell \rangle \equiv \frac{(\hat{a}^\dag)^{M-mn}(\hat{b}^\dag)^{kn+\ell}}{\sqrt{(M-mn)!(kn+\ell)!}}|Vac\rangle,\nonumber\\
\end{eqnarray} 
where $[\ldots]$ denotes  the integer part. The corresponding   $\beta^{(\mathcal{I})}_n$ in each invariant subspace $\cH_\mathcal{I}$  is  determined  from Eqs. (\ref{Psi_n}) and (\ref{model_1}):
\be
\beta^{(\mathcal{I})}_n = \left[\prod_{i=0}^{m-1}(M-mn-i) \right]\prod_{j=1}^k(kn+\ell+j).
\en{beta_1}
Since $N$ in  Eq. (\ref{model_1}) is   integer,    there is an  integer  $q$,  such that  $M = Nm + q$ and   $0 \leq q \leq m-1$. The    finite dimension of the subspace,   $\mathrm{dim}(\cH_\mathcal{I})=N+1$, is insured by $\beta^{(\mathcal{I})}_N =0$  in  Eq. (\ref{beta_1}), where at least one factor in the first product vanishes.
 
In the general multi-mode case of Eq. (\ref{models}),  the invariant subspaces are labeled by the composite index  $\mathcal{J} \equiv  (N,\ell_1,\ldots,\ell_S)$, where   $0\le \ell_s \le k_s-1$. The corresponding basis states are Fock states, analogous to Eq.~(\ref{model_1}). In this case, the  $\beta$-parameter reads
\be
\beta^{(\mathcal{J})}_n = \left[\prod_{i=0}^{m-1}(M-mn-i) \right]\prod_{s=1}^S\prod_{j=1}^{k_s}(k_sn+\ell_s+j).
\en{beta_2}

In  the optical applications  \cite{ClasskSq,GenSqueez} one typically has  $m=1$ in Eq. (\ref{models}),  where  the boson operator  $\hat{a}$  describes  the pump mode, whereas  the optical   modes labeled by $s$  correspond to the output (signal) modes.   For example,  the    $k$-photon down-conversion process into a single signal mode \cite{GenSqueez} is a special case of Eq. (\ref{models}) with $m=S=1$ and arbitrary $k\equiv k_1$. In this case,  the interaction Hamiltonian reads 
\be
\hat{H} = \hat{a}^\dag \hat{b}^k + \hat{a}(\hat{b}^\dag)^k\en{H_k}
 and the corresponding ladder coefficients in the invariant subspace are
\be
\beta^{(N)}_n = (N-n)\prod_{j=1}^k(kn+j).
\en{beta_k}

\section{Evolution   of arbitrary state }
\label{sec3}
 
We introduce the auxiliary state-number operator $ \hat{n}$, which   in  each invariant subspace $\mathcal{H}_N$  acts as  
\be
\hat{n}|\Psi_n\rangle = n |\Psi_n\rangle, \quad n=0,\ldots, N. 
\en{GTP3}
The   operators $\hat{n}$, $\hat{A}$, and $\hat{A}^\dag$   satisfy the  deformed  boson algebra: 
 \be
 [\hat{n},\hat{A}^\dag] = \hat{A}^\dag,\quad [\hat{n},\hat{A}] = -\hat{A},
 \en{nAAc}
since 
 \be
 \hat{A}\hat{A}^\dag=  \beta_{\hat{n}} ,\quad \hat{A}^\dag  \hat{A}=  \beta_{\hat{n}-1}.
\en{AAd}
Here $\beta_{\hat{n}}$ is obtained by replacing the scalar index   ``$n$"  in   $\beta_n$ by  the operator $\hat{n}$   Eq. (\ref{GTP3})  and we  set   $\beta_{-1}\equiv 0$.    From Eq. (\ref{nAAc})   one also obtains the  identity valid for an arbitrary scalar function $F(x)$:
 \be
F(\hat{n}) \hat{A}^\dag = \hat{A}^\dag F(\hat{n}+1).
\en{FA}
  
 We work in the interaction picture, where the Hamiltonian is given by Eq.~(\ref{H}) (recall that $[\hat{H}_0, \hat{H}] = 0$, with $\hat{H}_0$ the   free Hamiltonian). Consider the evolution of an arbitrary initial state within the subspace $\mathcal{H}_N$, using dimensionless variables such as the dimensionless time $\tau$. To this end, we analyze the evolution of each basis state in  Eqs.~(\ref{H_N})–(\ref{Psi_n}). To simplify the expressions, below we will  mainly employ the rescaled states
$(-i \hat{A}^\dag)^k |\Psi_0\rangle$, $k = 0, \ldots, N$,
and expand the unitary evolution operator in this basis as 
 \be
e^{-i\tau(\hat{A}+\hat{A}^\dag)}(-i \hat{A}^\dag)^k|\Psi_0\rangle= \sum_{n=0}^N \gamma_{n,k}(\tau) (-i\hat{A}^\dag)^n|\Psi_0\rangle.
\en{psit} 
The factor $(-i)^n$ is introduced to ensure that the expansion coefficients $\gamma_{n,k}(\tau)$ are real-valued functions.

\subsection{The states in the expansion of the evolution operator}
\label{sec3A}

Expanding the unitary evolution operator  $e^{-i\tau(\hat{A}+\hat{A}^\dag)}$ on the left-hand side of  Eq.~(\ref{psit}) generates states of the form
\be
|A_{m,k}\rangle\equiv (\hat{A} +\hat{A}^\dag)^m(\hat{A}^\dag)^k|\Psi_0\rangle.
\en{Amk}
 As shown in Appendix \ref{appA}, the state in Eq. (\ref{Amk}) can be expressed as
  \begin{eqnarray}
  \label {Amk_EX}
 |A_{m,k}\rangle &=& \sum_{l=0}^{L_{m,k}} a^{(l)}_{m,k}(\hat{A}^\dag)^{k+m-2l}|\Psi_0\rangle, \\
 L_{m,k}&\equiv& \mathrm{min}\left(m,\left[\frac{k+m}{2}\right]\right),\nonumber
 \end{eqnarray}
where  $[\ldots ]$  denotes   the integer part, and $a^{(l)}_{m,k}$ are scalar coefficients, which are combinations of the $\beta_n$ parameters defined in Eq.~(\ref{Psi_n}).

Let us derive the recursion relation for the scalars $a^{(l)}_{m,k}$. Multiplying Eq.~(\ref{Amk_EX}) on the left by $\hat{A}+\hat{A}^\dag$ and using Eq.~(\ref{AAd}) yields
\begin{eqnarray}
   |A_{m+1,k}\rangle &= & \left[ \sum_{l=0}^{L_{m,k}} a^{(l)}_{m,k}(\hat{A}^\dag)^{k+m+1-2l} \right. \nonumber \\
 & & \left. + \sum_{l=0}^{L_{m,k}}a^{(l)}_{m,k}  \beta_{\hat{n}} (\hat{A}^\dag)^{k+m-1-2l}  \right]|\Psi_0\rangle.\nonumber
 \end{eqnarray}

In the second sum, applying Eq.~(\ref{FA}) repeatedly with $F(\hat{n}) = \beta_{\hat{n}}$ and introducing a new summation index $l^\prime = l+1$ gives
\begin{eqnarray}
  \label {Am+1k}
&&   |A_{m+1,k}\rangle =  \left[ \sum_{l=0}^{L_{m,k}} a^{(l)}_{m,k}(\hat{A}^\dag)^{k+m+1-2l} \right.   \\
&& \left. + \sum_{l^\prime=1}^{L_{m,k}+1} \beta_{k+m+1-2l^\prime} a^{(l^\prime-1)}_{m,k}   (\hat{A}^\dag)^{k+m+1-2l^\prime} \right]|\Psi_0\rangle.\nonumber
 \end{eqnarray}
Since $\beta_{-1} = 0$ (and more generally $\beta_p = 0$ for $p<0$), we require $k+m+1 \ge 2 l^\prime$. Moreover, by definition in Eq.~(\ref{Amk_EX}), $a^{(l)}_{m,k} \equiv 0$ for $l  > m$. Hence, the upper limit $L_{m,k}+1$ in the second sum in Eq.~(\ref{Am+1k}) can be replaced with $L_{m+1,k}$. The same replacement can be performed in the first sum if we postulate
\be
a^{(l)}_{m,k} \equiv 0, \quad   l\notin [0,L_{m,k}].
\en{C_alm}
Finally, combining the two sums in Eq.~(\ref{Am+1k}) and comparing with Eq.~(\ref{Amk_EX}) for $m+1$ (then replacing $m +1\to m$) yields the desired recursion relation:
\be
a^{(l)}_{m,k} = a^{(l)}_{m-1,k}+ \beta_{k+m-2l} a^{(l-1)}_{m-1,k}.
\en{Ralm}

The boundary  values of $a^{(l)}_{m,k}$ for $l=0$ and $l=m$ correspond to the simplest cases of the recursion  in Eq.~(\ref{Ralm}).  Setting $l=0$ in Eq. (\ref{Ralm}) and taking into account 
Eq.~(\ref{C_alm}) we obtain
\be
a^{(0)}_{m,k} = a^{(0)}_{m-1,k} = \ldots = a^{(0)}_{0,k}=1. 
\en{a0m}
Setting $l=m$ for $m\le k$  in Eq. (\ref{Ralm}) and using $ a^{(0)}_{0,k}=1$ we get:
\be
a^{(m)}_{m,k} = \beta_{k-m}a^{(m-1)}_{m-1,k} = \ldots = \prod_{s=k-m}^{k-1}\beta_s, \quad \forall m\le k.
\en{amm} 
 For $m>k$,  Eq.~(\ref{amm}) yields zero, since $\beta_p = 0$ for $p<0$. This agrees with the condition on $a^{(l)}_{m,k}$  in 
 Eq.~(\ref{C_alm}),  which allows non-zero values only for  $l \le L_{m,k} =[ (k+m)/2]<m$ in this case.

\subsection{Solving the recursion; the   $g$-factors }
\label{sec3B}

It proves convenient for what follows to reindex the scalars  $a^{(l)}_{m,k}$ of Eq. (\ref{Amk_EX}) by   introducing new ones $g^{(l)}_{n,k}$, defined as 
\be
g^{(l)}_{k+m-2l,k} \equiv a^{(l)}_{m,k}. 
\en{gln}
Then, Eq. (\ref{C_alm})  together with the relation   $m=n-k+2l$ can be cast in the form
\be
g^{(l)}_{n,k}\equiv  0, \quad l<0\cup n<0 \cup 2l<k-n. 
\en{C_gln}
Let us also rewrite the recursion   in Eq. (\ref{Ralm}) in terms of  $g^{(l)}_{n,k}$. We get
\be
g^{(l)}_{n,k} = g^{(l)}_{n-1,k} +\beta_n g^{(l-1)}_{n+1,k}.
\en{Rgln} 
 The boundary conditions in Eqs. (\ref{a0m})-(\ref{amm}) become
\begin{eqnarray}
\label{Bgln1}
&& g^{(0)}_{n,k}= 1,\quad n\ge k,\\
&& g^{(k-n)}_{n,k} = \prod_{s=n}^{k-1}\beta_s,\quad  n\le k-1.  
\label{Bgln2}	
\end{eqnarray}
If the empty product is set to unity,   then the two conditions in Eq. (\ref{Rgln}) coincide  for $n=k$.  We will adopt this convention below.

Now,  it is easy to show that   $g^{(l)}_{n,k}=0 $ for $l<k-n$, i.e.,     the  (nonzero) boundary values of $g^{(l)}_{n,k}$ for $n<k$ are  given by Eq. (\ref{Bgln2}). To this goal let us  rewrite Eq. (\ref{Rgln}) for $n+1$ in the form suitable for induction:
\be
g^{(l)}_{n,k} = g^{(l)}_{n+1,k} - \beta_{n+1}g^{(l-1)}_{n+2,k}.
\en{Rgln_I}
For  $k-n-1\ge 1$, i.e., $n+1\le k-1$ by setting $l = k-n-1$  in Eq. (\ref{Rgln_I})  and  using Eq. (\ref{Bgln2})   we  obtain:
\begin{eqnarray}
\label{Zerog1}
&& g^{(k-n-1)}_{n,k} = g^{(k-[n+1])}_{n+1,k} - \beta_{n+1}  g^{(k-[n+2])}_{n+2,k} \nonumber \\
&&= \prod_{s=n+1}^{k-1}\beta_s - \beta_{n+1} \prod_{s=n+2}^{k-1}\beta_s=0.
\end{eqnarray}
In its turn, by setting  $l=n-k-2$ in  Eq. (\ref{Rgln_I}) and using the result in Eq. (\ref{Zerog1})  for $n+1$ and for $n+2$ we get:
\be
g^{(k-n-2)}_{n,k} = g^{(k-[n+1]-1)}_{n+1,k} - \beta_{n+1}g^{(k-[n+2]-1)}_{n+2,k}=0. 
\en{Zerog2}
In this way, by decreasing $l$ one step at a time and repeatedly applying  Eq. (\ref{Rgln_I}),  one obtains  on the right-hand side only the zero values  derived in the preceding steps. This shows  that $g^{(l)}_{n,k}=0$ for  all $l< k-n$.   Combining  this result with the conditions from  Eq. (\ref{C_gln}) we have:
\be
g^{(l)}_{n,k}=0, \quad   l < l_0(k,n)\equiv \mathrm{max}(k-n,0) \cup n<0. 
\en{gln=0}
 
Now,   let us  solve the two-dimensional recursion in Eq.~(\ref{Rgln})  under the boundary conditions in Eqs.~(\ref{Bgln1})-(\ref{Bgln2}). 
To this goal we iterate    Eq.~(\ref{Rgln}) with respect to  the index ``$n$"   in  the  decreasing direction,  using     Eq.~(\ref{Rgln}) itself on  the first term on its  right-hand side: 
\begin{eqnarray*}
\label{RglmIT}
 g^{(l)}_{n,k} &= & g^{(l)}_{n-1,k} +\beta_n g^{(l-1)}_{n+1,k}\\
&=&  g^{(l)}_{n-2,k} +\beta_{n-1} g^{(l-1)}_{n,k}+\beta_n g^{(l-1)}_{n+1,k}\\
& = & g^{(l)}_{n-3,k} + \beta_{n-2} g^{(l-1)}_{n-1,k} + \beta_{n-1} g^{(l-1)}_{n,k}+\beta_n g^{(l-1)}_{n+1,k}\\
&=&\ldots.
\end{eqnarray*}
Continuing this procedure, the first term eventually reaches the index at which it vanishes   by Eq. (\ref{gln=0}), namely  $g^{(l)}_{k-l-1,k}=0$.  Hence the iteration yields a one-dimensional recursion in the ``$l$"-index:
\be
g^{(l)}_{n,k} = \sum_{s=k-l}^n \beta_s g^{(l-1)}_{s+1,k}.
\en{Rgln_1}
 Iteration of  the one-dimensional recursion in  Eq. (\ref{Rgln_1}) produces the nested-sum representation
\be
g^{(l)}_{n,k} = \sum_{s_1=k-l}^n\beta_{s_1}\sum_{s_2 = k-l+1}^{s_1+1}\beta_{s_2}\ldots\sum_{s_l=k-1}^{s_{l-1}+1}\beta_{s_l},
\en{gln_E}
where we have taken into account that $g^{(0)}_{s_l+1,k} =1$ by Eq. (\ref{Bgln1}). The expression  in Eq. (\ref{gln_E}) reduces to the previously obtained formula for $k=0$  \cite{My1}, in which case the lower limits may be taken as $s_j\ge 0$, since $\beta_{p}=0$ for $p<0$. 

\subsection{The  $g$-factors as powers of a Hessenberg matrix}
\label{sec3C}

For   the special initial state $|\Psi_0\rangle$ ($k=0$)  the  nested sum of Eq. (\ref{gln_E}) was rewritten in Ref.  \cite{My1} using  some auxiliary Hessenberg matrix.  The same approach generalizes to an arbitrary basis state $|\Psi_k\rangle$ ($k\ge 0$).  Introduce the   orthonormal basis  of column-vectors   
\be 
  |e_0\rangle \equiv \left(\begin{matrix}1\\0\\ \vdots\\0 \end{matrix}\right),  \ldots,   |e_N\rangle \equiv \left(\begin{matrix}0\\ \vdots\\0\\1  \end{matrix}\right) 
 \en{vect}
 in an auxiliary linear space of dimension $N+1$,  and denote the corresponding row vectors in the dual (conjugate) space by $\langle e_0|, \ldots, \langle e_N|$. Let matrix $\mathbf{B}$ be as follows 
\be
 \mathbf{B} = \sum_{n=0}^N|e_n\rangle\sum_{s=0}^n\beta_s\langle e_{s+1}| .
\en{matB}
The   matrix  $\mathbf{B}$  of  rank $N$ belongs to the class of  lower Hessenberg matrices (in contrast to the lower triangular matrix, the first super-diagonal can be nonzero also). Powers of such matrices are easy to compute numerically  \cite{Hessen}.  The $l$th power of  $\mathbf{B}$ admits the representation
\be
\mathbf{B}^l = \sum_{n=0}^N|e_n\rangle\sum_{s_1=0}^n\beta_{s_1}\sum_{s_2=0}^{s_1+1}\beta_{s_2}\ldots\sum_{s_l=0}^{s_{l-1}+1}\beta_{s_l}\langle e_{s_l+1}|.
\en{matB2l}
To recover the nested sums of Eq. (\ref{gln_E}) one evaluates the matrix element
\be
g^{(l)}_{n,k} = \langle e_n| \mathbf{B}^l \sum_{m=k}^N|e_m\rangle.
\en{gln_B}
Indeed,  Eq. (\ref{matB2l}) implies  $s_l\ge k-1$,  thus $s_{l-1}\ge s_l-1=k-2$ from the  upper limit of the innermost  nested sum.  By  utilizing repeatedly   the upper limits  in Eq.~(\ref{matB2l}),       $s_{j-1}\ge s_j -1$, we obtain   $s_j\ge k-l+j-1$, i.e.,  the lower limits  in Eq. (\ref{gln_E}). 

Moreover, introducing the  lower triangular  matrix  $\mathbf{T}$:
\be 
 \mathbf{T} =\sum_{n=0}^N \sum_{p=0}^{N-n}  |e_{n+p}\rangle \langle e_{n}|
\en{T}
(the elements in the lower triangular part   are all  equal to  $1$), we get an equivalent expression  
\be
g^{(l)}_{n,k} = \langle e_n| \mathbf{B}^l \mathbf{T} |e_k\rangle.
\en{gln_BT}
 
 \subsection{The state evolution: main results}
 \label{sec3D}
 
The  above analysis  can be summarized as follows. 
\begin{theorem}
The   state \mbox{$|A_{m,k}\rangle \equiv (\hat{A} +\hat{A}^\dag)^m(\hat{A}^\dag)^k|\Psi_0\rangle$} is given by 
\be
|A_{m,k}\rangle =  \sum_{l=0}^{L_{m,k}} g^{(l)}_{k+m-2l,k}(\hat{A}^\dag)^{k+m-2l}|\Psi_0\rangle,
\en{m-state}
where  $L_{m,k}= \mathrm{min}\left(m,\left[\frac{k+m}{2}\right]\right)$ (with $[\ldots]$ denoting the integer part) and   the scalars  $g^{(l)}_{n,k}$  are  given by Eq. (\ref{gln_E}) or, alternatively,  by Eqs. (\ref{gln_B})-(\ref{gln_BT}).
\end{theorem}
  \begin{corollary}
The time evolution of the rescaled basis state $(-i \hat{A}^\dag)^k |\Psi_0\rangle$  is as follows:
\begin{eqnarray}
\label{psi-state}
&&e^{-i\tau(\hat{A} +\hat{A}^\dag)} (-i\hat{A}^\dag)^k |\Psi_0\rangle=\sum_{n=0}^N \gamma_{n,k}(\tau) (-i\hat{A}^\dag)^n|\Psi_0\rangle,\nonumber\\
&& \gamma_{n,k}(\tau) =  \sum_{l=l_0(k,n)}^\infty \frac{(-1)^l \tau^{n-k+2l}}{(n-k+ 2l)!} g^{(l)}_{n,k},
\end{eqnarray}
where $l_0(k,n)=\mathrm{max}(k-n,0)$. In other words,  the quantum amplitude  $\gamma_{n,k}(\tau)$  is a power series expansion in $\tau$ with  the  coefficients $g^{(l)}_{n,k}$ being given by Eq. (\ref{gln_E}) or, alternatively,  by Eqs. (\ref{gln_B})-(\ref{gln_BT}).
\end{corollary}
\noindent\textit{Proof.--}  
To prove Eq. (\ref{psi-state}), we substitute the result of Theorem 1 into the power series expansion of the evolution operator,  interchange the order of the summation and introduce a new    index  $n\equiv m+k-2l$:
\begin{eqnarray*}
\label{proofC}
&& e^{-i\tau(\hat{A} +\hat{A}^\dag)} (-i\hat{A}^\dag)^k|\Psi_0\rangle = \sum_{m=0}^\infty \frac{(-i\tau)^m}{m!}(-i)^k|A_{m,k}\rangle\\
&& = \sum_{n=0}^N \left(\sum_{l=l_0(k,n)}^\infty\frac{(-1)^l\tau^{n-k+2l}}{(n-k+2l)!}g^{(l)}_{n,k}\right)(-i\hat{A}^\dag)^n|\Psi_0\rangle,
\end{eqnarray*}
where   we have taken into account that the  interval $0\le l\le L_{m,k} = L_{n-k+2l,k}$ is equivalent to two conditions: $2l \le n+2l$ and $l\le n-k+2l$, the first  giving   $n\ge 0$, while the second \mbox{$l\ge \mathrm{max}(0,k-n)=l_0(k,n)$.}
The sum  inside the parentheses is the expression for  $\gamma_{n,k}(\tau)$  given in Eq. (\ref{psi-state}). Q.E.D.

 Some observations are in order.   From the nested-sum expression in  Eq. (\ref{gln_E})    one  can infer by induction that   $s_j \le n+j-1$, thus:
\be
g^{(l)}_{n,k} \le      \max_{0\le p\le l-1}\left( \sum_{s = k-l+p}^{n+p}\beta_s\right)^l\le \left( \sum_{s = k-l}^{n+l-1}\beta_s\right)^l.
\en{Up_bound_g} 
Since the sum over $\beta_n$ is finite, Eq. (\ref{Up_bound_g})  ensures that the power series in Eq. (\ref{psi-state})  converges regardless of the values of the $\beta$-factors (the series of the absolute values  is bounded by a uniformly convergent series).  Hence,   $\gamma_{n,k}(\tau)$ in Eq. (\ref{psi-state})  is a holomorphic  function in the complex plane of $\tau$.

One can  give an alternative  expression for $\gamma_{n,k}$ as follows:
\begin{eqnarray}
\label{gam_n,k}
\gamma_{n,k}(\tau) &=& \sum_{l=0}^\infty \frac{(-1)^{l+l_0(k,n)} \tau^{|n-k|+2l}}{(|n-k|+ 2l)!} \langle e_n| \mathbf{B}^{l+l_0(k,n)}\mathbf{T} |e_k\rangle\nonumber\\
&=&  \sum_{l=0}^\infty \frac{(-1)^l \tau^{n-k+2l}}{(n-k+ 2l)!} \langle e_n| \mathbf{B}^l \mathbf{T} |e_k\rangle,
 \end{eqnarray}
where in the last sum  the actual summation  starts with  $l\ge l_0(k,n)$, since   $g^{(l)}_{n,k}$   are all   zero for $l<l_0(k,n)$ (the matrix average is zero, as can be easily verified). 
If we introduce a series of matrix functions of $\tau$:
\be
\mathbf{\Gamma}_m\equiv  \sum_{l=0}^\infty \frac{(-1)^l \tau^{m+2l}}{(m+ 2l)!}   \mathbf{B}^l \mathbf{T},
\en{gamMat}
then (with the above observation on the lower limit of the summation)
\be
\gamma_{n,k}(\tau) = \langle e_n| \mathbf{\Gamma}_{n-k}(\tau)|e_k\rangle = \frac{\mathrm{d}^k}{\mathrm{d}\tau^k}\langle e_n| \mathbf{\Gamma}_{n}(\tau)|e_k\rangle. 
\en{gam_derk}

The  results of  Theorem 1 and Corollary 1 reduce to the previously derived  results    \cite{My1} for  $k=0$.  For instance, it was verified in Ref.  \cite{My1} by direct substitution and  using  the recursion  Eq. (\ref{Rgln})  that for $k=0$ Eq. (\ref{psi-state}) solves the evolution equation.  For $k\ge 0$   the verification by direct substitution in Ref.  \cite{My1} also works since the recursion is the same. The initial condition is guaranteed by the observation on the actual summation below Eq. (\ref{gam_n,k}).  

From Eqs. (\ref{Psi_n}) and (\ref{psi-state})-(\ref{gam_n,k}) one  immediately obtains the following. 
\begin{corollary}
The average of the evolution operator with the Hamiltonian of Eqs. (\ref{H})-(\ref{GTP2})  on two arbitrary basis states from Eq. (\ref{Psi_n}) reads:
\begin{eqnarray}
\label{evol_n,k}
\langle \Psi_n| e^{-i\tau(\hat{A}+\hat{A}^\dag)}|\Psi_k\rangle = \left[ \frac{\prod_{p=0}^{n-1}\beta_p}{\prod_{q=0}^{k-1}\beta_q}\right]^\frac{1}{2}\nonumber\\
\times  \sum_{l=l_0(k,n)}^\infty \frac{ (-i\tau)^{n-k+2l}}{(n-k+ 2l)!} \langle e_n| \mathbf{B}^l \mathbf{T} |e_k\rangle,
\end{eqnarray}
where the   orthonormal vectors $|e_m\rangle$, $m=0,\ldots, N$ span an auxiliary vector space (isomorphic  to the considered subspace $\cH_N$ of the Hilbert space),    the lower Hessenberg matrix  $\mathbf{B}$ is defined  in   Eq. (\ref{matB}) and  the  lower-triangular matrix  $\mathbf{T}$   in  Eq.  (\ref{T}).  Here $l_0(k,n) = \mathrm{max}(0,k-n)$, i.e.,  for $n<k$ the nonzero terms   have $l\ge k-n$. 
Alternatively, by using Eq. (\ref{gam_n,k}), Eq. (\ref{evol_n,k}) can be  rewritten as 
\begin{eqnarray}
\label{Ead1}
 \langle \Psi_n| e^{-i\tau(\hat{A}+\hat{A}^\dag)}|\Psi_k\rangle =(-i)^{n-k} \left[ \frac{\prod_{p=0}^{n-1}\beta_p}{\prod_{q=0}^{k-1}\beta_q}\right]^\frac{1}{2}\gamma_{n,k}(\tau)\nonumber\\
 \end{eqnarray}
 with $\gamma_{n,k}(\tau)$ given by Eq. (\ref{psi-state}).
\end{corollary}

\section{The energy spectrum }
\label{sec4}
One   approach to determining the   energy spectrum of a solvable model is to use the algebraic or group-theoretic methods \cite{GroupMeth,LieAlg3Boson,GroupMeth2,LieAlgGen1,LieAlgGen2}. Here an alternative   approach is developed.   The characteristic polynomial  for the energy eigenvalues is derived from  a  recursion, where the coefficients are expressed as nested sums. Alternatively, the characteristic polynomial can be represented by a continued fraction.   The amplitudes of the eigenstates are  given in  several equivalent forms:  as continued fractions (or, equivalently, as  a product of the  M\"obius  group  matrices in two-dimensional projective space),   as nested sums, and  as the  principal minors of a Jacobi matrix.

Consider the eigenvalue problem  in an invariant subspace $\cH_N$ for    the Hamiltonian of Eq. (\ref{H}).  Expanding  eigenstates in the basis of  Eq. (\ref{Psi_n}),
\be
|\Psi(\tau)\rangle  = e^{-i\lambda \tau} \sum_{n=0}^N\psi_n|\Psi_n\rangle,
\en{4Eq1}
we obtain the stationary  Schr\"odinger equation in the following form
\be
\lambda \psi_n = \sqrt{\beta_{n-1}}\psi_{n-1} + \sqrt{\beta_n}\psi_{n+1},
\en{4Eq2}
where  $  \psi_{-1} = \psi_{N+1}=0$ (recall that $\beta_{-1} = \beta_{N}=0$).  Observe also that, for an  eigenstate, $\psi_0\ne0$ and $\psi_N\ne 0$.  Eq. (\ref{4Eq2}) remains  invariant under the  transformation: $\lambda\to -\lambda$ and $\psi_n\to (-1)^n\psi_n$. Hence,  the nonzero eigenvalues come in pairs $\pm \lambda$.

\subsection{The spectrum  via   continued fractions}
\label{sec4A}

Consider  the rescaled  ratio of two nearest quantum amplitudes
\be
R_n\equiv \frac{\sqrt{\beta_n} \psi_n}{\psi_{n+1}}, \quad n=0,\ldots, N-1,
\en{Rn_psi}
or its inverse. The boundary conditions on $\psi_n$   require that $R_{-1}=0$ and $R_{N}=\infty$.
  Eq. (\ref{4Eq2}) tells us that  the ratio $R_n$ satisfies  a simple recursion, which can be cast in two equivalent forms: 
\be
R_n = \frac{\beta_n}{\lambda - R_{n-1}},\quad \frac{\beta_n}{R_n} = \frac{\beta_n}{\lambda - \frac{\beta_{n+1}}{R_{n+1}}}.
\en{4Eq4}
The boundary conditions are 
\be
   \frac{\beta_0}{R_0} =   R_{N-1} = \lambda.
\en{4Eq5}
 One of the relations  in Eq. (\ref{4Eq5})  gives   $R_n$ at either $n=0$ or $n = N-1$, while the other --  the  characteristic equation for the energy eigenvalues $\lambda$.   The  existence of  two equivalent forms for the characteristic equation  follows from the   finite dimension of the invariant subspace $\cH_N$ and the inversion  symmetry of the  row and column  indices  in the  matrix formulation of the  eigenvalue problem in  $\cH_N$, i.e., $(\beta_0,\beta_1,\ldots, \beta_{N-1}) \rightarrow (\beta_{N-1},\beta_{N-2},\ldots, \beta_{0})$ (the characteristic equation is the  determinant of a Jacobi matrix, see  Eqs. (\ref{matIJ})-(\ref{EigProbJ}) below).

The solution to  Eqs. (\ref{4Eq4})-(\ref{4Eq5})  can be conveniently cast  in the form of two  continued fractions:
\be
R_n =  \cfrac{\beta_n}{\lambda - \cfrac{\beta_{n-1}}{\lambda - \cfrac{\beta_{n-2}}{  \cfrac{ \ddots
}{\lambda - \cfrac{\beta_{0}}{\lambda}}}}},\quad \frac{\beta_n}{R_n} =  \cfrac{\beta_n}{\lambda - \cfrac{\beta_{n+1}}{\lambda - \cfrac{\beta_{n+2}}{  \cfrac{ \ddots
}{\lambda - \cfrac{\beta_{N-1}}{\lambda}}}}},\qquad
\en{ContFr}
where one of the two  boundary conditions in Eq. (\ref{4Eq5}) specifies  the  lowest  denominator   in the respective  continued fraction and the other -- the  characteristic equation for $\lambda$.

 One can use the  M\"obius group  to represent a  continued fraction   \cite{BookCF}. The denominators in the continued  fractions of Eq. (\ref{ContFr})   result from  the application of the  following M\"obius transformation:
\be
  z\to \mathbf{M}_s(z) \equiv \frac{-\beta_s}{\lambda +z}.
\en{Mobs} 
Therefore, the continued-fraction representations  in Eq. (\ref{ContFr}) become:
\begin{eqnarray}
\label{MContFr1}
-R_n &=& \mathbf{M}_n\left(\mathbf{M}_{n-1}\left(\ldots \mathbf{M}_0(0)\right)\ldots \right), \\
-\frac{\beta_n}{R_n} &=&  \mathbf{M}_n\left(\mathbf{M}_{n+1}\left(\ldots \mathbf{M}_{N-1}(0)\right)\ldots \right).
\label{MContFr2}
\end{eqnarray}
Introduce the following matrix
\be
\mathbf{M}_s \equiv \left(\begin{matrix} 0 & -\beta_s \\
1 & \lambda \end{matrix}\right).
\en{Ms}
There is a simple  relation between  the  product of matrices $\mathbf{M}_s$      and   the nested  M\"obius action of Eqs. (\ref{Mobs}), (\ref{MContFr1})-(\ref{MContFr2}). We have 
\be
 \left(\begin{matrix} p_s\\ q_s \end{matrix}\right) \equiv    \mathbf{M}_s\left(\begin{matrix} cz\\ c\end{matrix}\right), \quad \mathbf{M}_s (z) = \frac{p_s}{q_s},\quad  \forall c\ne 0 .
\en{IdMs} 
Hence,  instead of the series of nested M\"obius transformations, one may alternatively   compute  the product of the respective M\"obius matrices Eq. (\ref{Ms})   in Eqs. (\ref{MContFr1})-(\ref{MContFr2}) and use Eq.~(\ref{IdMs}). For instance,
\be 
  R_n = -\frac{P_n}{Q_n},\quad  \left(\begin{matrix} P_n\\ Q_n \end{matrix}\right) \equiv\mathbf{M}_n\mathbf{M}_{n-1}\ldots \mathbf{M}_0  \left(\begin{matrix} 0\\ 1 \end{matrix}\right).
\en{RnMob}
 Moreover,  the   identity
\be
(1,0) \mathbf{M}_n = (0,-\beta_n)
\en{IdM2}
relates the first and the second row in the  products of   $n$ and $n-1$   M\"obius  matrices in  Eq. (\ref{RnMob}). Therefore, we can  write the matrix product      as follows
\be
 \mathbf{M}_n\mathbf{M}_{n-1}\ldots \mathbf{M}_0 = \left(\begin{matrix}   -\beta_n & 0 \\ 0 & 1 \end{matrix}\right)\left(\begin{matrix}  X_{n}& Y_{n} \\ 
X_{n+1}& Y_{n+1} \end{matrix}\right), 
\en{Mprod1}
 where by setting  $n=0$ we get: $X_{0} = 0$, $X_1 = 1$ and $Y_{0} = 1$, $Y_1 = \lambda$.  Then
 \be 
   R_n = \frac{\beta_n Y_{n}}{Y_{n+1}}.
 \en{Rn_Y}

 Now, in order  to find the explicit form of  the characteristic polynomial for  the energy eigenvalues $\lambda$ in Eq.~(\ref{4Eq5}), we will find a recursion for the quantities $Y_n$ in Eqs.~(\ref{Mprod1})-(\ref{Rn_Y}).  Evaluating  the product in Eq. (\ref{Mprod1})  by  using Eq.~(\ref{Mprod1}) for $n-1$ and, alternatively, by  multiplying by $\mathbf{M}_{n}$ from the left, we obtain: 
 \begin{eqnarray*}
&&\mathbf{M}_n\mathbf{M}_{n-1}\ldots \mathbf{M}_0   = \left(\begin{matrix}   0 & -\beta_n \\ 1 & \lambda \end{matrix}\right) \mathbf{M}_{n-1}\ldots \mathbf{M}_0 \\
&& = \left(\begin{matrix}   0 & -\beta_n \\ -\beta_{n-1}& \lambda \end{matrix}\right)
\left(\begin{matrix}   X_{n-1} & Y_{n-1} \\ X_{n} & Y_{n}\end{matrix}\right)\\
&&=\left(\begin{matrix}   -\beta_n & 0 \\ 0 & 1 \end{matrix}\right)\left(\begin{matrix}   X_{n} & Y_{n} \\ X_{n+1} & Y_{n+1}\end{matrix}\right).
\end{eqnarray*}
Comparing  the  matrix entry  at   position  $(2,2)$ on the both  sides of the above identity, we get the   recursion
\be 
Y_{n+1} = \lambda Y_{n} - \beta_{n-1}Y_{n-1},
\en{RecY}
with the initial values:    $Y_0=1$  and $Y_{-1}\equiv 0$ ($Y_{-1}$, which  appears for $n=0$, is   multiplied by $\beta_{-1}=0$). Observe that the characteristic equation $R_{N-1}(\lambda)= \lambda$ Eq. (\ref{4Eq5}) in terms of $Y_n$ reads:
\be
Y_{N+1}(\lambda) \equiv \lambda Y_{N}(\lambda)  -\beta_{N-1}Y_{N-1}(\lambda) = 0,
\en{CharY}
where we have  introduced the quantity $Y_{N+1}$, which satisfies the  recursion of Eq.~(\ref{RecY}), but is absent from the  product representation  in Eqs. (\ref{Mprod1})-(\ref{Rn_Y}),   where we have $n\le N-1$. 

One final  observation is in order. From Eqs. (\ref{4Eq2})-(\ref{Rn_psi}) and (\ref{Rn_Y})-(\ref{RecY})  one can easily establish  the following  relation between $\psi_n$ and $Y_{n-1}$:
\be
\psi_n \equiv  \psi_0 \left(\prod_{s=0}^{n-1}\beta_s\right)^{-\frac12} Y_{n}.
\en{psinYn-1}
 Indeed,  if we use  Eq. (\ref{psinYn-1}) as an alternative definition of $Y_n$, then from  Eq. (\ref{4Eq2})   it follows that the quantity $Y_n$ satisfies the recursion of Eq. (\ref{RecY}) and the  initial  conditions: $Y_{-1}=0$ and $Y_{0}=1$ (recall that the empty product is   equal to  $1$).   
 
 The    characteristic equation for the  eigenvalues  of   Eq.~(\ref{4Eq2}) reads: $\psi_{N+1}(\lambda) =0$, where $\psi_n(\lambda)$ is the general  solution.   We have  shown   that it also reads  $Y_{N+1}(\lambda)=0$ Eq.~(\ref{CharY}). Observe that the relation between the two forms by  Eq.~(\ref{psinYn-1})   for $n=N+1$ involves the singular first factor  on the right-hand side   (recall that $\beta_N=0$).

\subsection{The characteristic equation  for the energy eigenvalues}
\label{sec4B}

The recursion in Eq. (\ref{RecY})   defines    a sequence of   polynomials in  $\lambda$:    $Y_n =Y_n(\lambda)$. It  can be easily solved   in a  similar way as  the recursion in Eq. (\ref{Rgln_I}).  We have  the following result. 
\begin{theorem}
The   solution to the recursion  in Eq. (\ref{RecY})  reads:
  \begin{eqnarray}
 \label{SolRecY}
&&  Y_{n}(\lambda) = \sum_{l=0}^{\left[ \frac{n}{2}\right]}(-1)^l\mathcal{G}^{(l)}_{n-1}  \lambda^{n-2l},\\
 && \mathcal{G}^{(l)}_n\equiv \sum_{s_1= 0}^{n-2l+1}\beta_{s_1} \sum_{s_2= s_1+2}^{n-2l+3}\beta_{s_2}\ldots  \sum_{s_l= s_{l-1}+2}^{n-1}\beta_{s_l}, \quad l\ge 1,   \nonumber
 \end{eqnarray}
and  $\mathcal{G}^{(0)}_n\equiv1$. 
 \end{theorem}

   \noindent{\textit{Proof.--} We will  use of the following recursive identity (for $l\ge 1$):
 \begin{eqnarray}
 \label{IdGln}
   \mathcal{G}^{(l)}_n =  \mathcal{G}^{(l)}_{n-1} + \beta_{n-1} \mathcal{G}^{(l-1)}_{n-2}.
   \end{eqnarray}
   The recursion in Eq. (\ref{IdGln}) easily follows from another equivalent expression for $\mathcal{G}^{(l)}_n$.   The   index in the $j$th nested sum  in Eq. (\ref{SolRecY}) satisfies \mbox{$2(j-1)\le s_j  \le  s_{j+1}-2$},   $j=1,\ldots, l-1$.  Hence,  the nested sums can be arranged  in the inverse order, i.e., from $l$ to $1$,  as follows
\be
 \mathcal{G}^{(l)}_n =  \sum_{s_l= 2(l-1)}^{n-1}\beta_{s_l} \sum_{s_{l-1}= 2(l-2)}^{s_l-2}\beta_{s_{l-1}}\ldots  \sum_{s_1= 0}^{s_2-2}\beta_{s_1}.
   \en{Gln_In}
Separating the term with $\beta_{n-1}$ in  the first sum  in   Eq. (\ref{Gln_In}),
\begin{eqnarray*}
 &&\mathcal{G}^{(l)}_n = \sum_{s_l= 2(l-1)}^{n-2}\beta_{s_l} \sum_{s_{l-1}= 2(l-2)}^{s_l-2}\beta_{s_{l-1}}\ldots  \sum_{s_1= 0}^{s_2-2}\beta_{s_1}\\
 &&+\beta_{n-1} \sum_{s_{l-1}= 2(l-2)}^{n-3}\beta_{s_{l-1}} \sum_{s_{l-2}= 2(l-3)}^{s_{l-1}-2}\beta_{s_{l-2}}\ldots  \sum_{s_1= 0}^{s_2-2}\beta_{s_1},
  \end{eqnarray*} 
 and replacing the  summation index  in the last inner  sum by $l^\prime\equiv   l-1$, we arrive at the identity in Eq. (\ref{IdGln}). 
 
 Now let us  prove that the polynomial of degree $n+1$ on the right hand side of  Eq. (\ref{SolRecY}) solves the recursion in Eq. (\ref{RecY}).   First of all, Eq.~(\ref{SolRecY})  reproduces the  first two polynomials: $Y_{0}=1$ and $Y_{1} = \lambda$  (for $n=0$ and $n=1$ there is only the term with $l=0$).  Now  consider the expression for  
 $Y_{n+1}(\lambda)$   provided by  Eq. (\ref{SolRecY}). By  separating the term with $l=0$ and using the identity (\ref{IdGln}),  we obtain:
 \begin{eqnarray*}
 &&Y_{n+1}(\lambda) = \lambda^{n+1} + \lambda  \sum_{l=1}^{\left[ \frac{n+1}{2}\right]}(-1)^l\mathcal{G}^{(l)}_{n-1}  \lambda^{n-2l}\\
 && + \beta_{n-1} \sum_{l=1}^{\left[ \frac{n+1}{2}\right]}(-1)^l\mathcal{G}^{(l-1)}_{n-2}  \lambda^{n+1-2l}\\
 && = \lambda  \sum_{l=0}^{\left[ \frac{n+1}{2}\right]}(-1)^l\mathcal{G}^{(l)}_{n-1}  \lambda^{(n-1)+1-2l}\\
 && - \beta_{n-1} \sum_{l^\prime=0}^{\left[ \frac{n+1}{2}\right]-1}(-1)^{l^\prime}\mathcal{G}^{(l^\prime)}_{n-2}  \lambda^{(n-2)+1-2l^\prime},
 \end{eqnarray*}
 where  we have set $l = l^\prime+1$ in the last sum. Let us  verify that the upper limits in  the above sums conform with  Eq.~(\ref{SolRecY}), thus giving $Y_{n}(\lambda)$ 
 and $Y_{n-1}(\lambda)$, respectively. 
In the first sum, by   virtue of Eq.~(\ref{Gln_In}),  the nonzero  values  of  $\mathcal{G}^{(l)}_{n-1}$  have   $2(l-1)\le (n-1)-1$,   thus we can replace the upper limit with $l\le \left[\frac{n}{2}\right]$. By a similar  argument, in the second sum we have $2(l-1)\le (n-2)-1$, i.e., the upper limit can be set to  $l\le \left[\frac{n-1}{2}\right]$. Thus we have arrived at the recursion in Eq. (\ref{RecY}). Q.E.D.
  
In order to obtain the rescaled quantum amplitudes of the energy eigenstates from  Eqs. (\ref{psinYn-1})-(\ref{SolRecY}), the  parameter $\lambda$  must be set to be one of the energy eigenvalues   $\lambda_j$, $j=1,\ldots, N+1$, of the stationary Schr\"odinger equation  (\ref{4Eq2}).  The characteristic equation is  $Y_{N+1}(\lambda) =0$ Eq. (\ref{CharY}), with $Y_{N+1}(\lambda)$ given by  Eq. (\ref{SolRecY}). We get the following. 
\begin{corollary} 
The characteristic  polynomial for the energy eigenvalues  $\lambda_j$, $j=1,\ldots, N+1$ in the invariant subspace $\cH_N$ reads
\be
 \sum_{l=0}^{\left[ \frac{N+1}{2}\right]}(-1)^l\mathcal{G}^{(l)}_N \lambda^{N+1-2l} = 0,
\en{YN}
where the coefficients $\mathcal{G}^{(l)}_N$ are defined by  Eqs. (\ref{SolRecY})-(\ref{Gln_In}). 
\end{corollary}
We have already found that the energy eigenvalues come in pairs $\pm \lambda$, except for the zero eigenvalue. Eq. (\ref{YN})  tells us that the (simple)   zero eigenvalue appears in all  odd-dimensional invariant subspaces $\cH_N$, i.e., when $N+1$ is an odd integer.

 \subsection{Amplitudes of the eigenstates as  Jacobi  determinants }
 \label{sec4C}
 
 There is an alternative expression  for the quantum amplitudes $\psi_n$  of an eigenstate by the  principal minors of  the associated Jacobi matrix. Consider  the sequence of  matrices for $1\le n \le N+1$:
 \begin{eqnarray}
&&   \mathbf{I}_n\equiv \sum_{s=0}^{n-1} |e_s\rangle\langle e_s|, \nonumber\\
&&\mathbf{J}_n \equiv  \sum_{s=0}^{n-2} \sqrt{\beta_s}\left(  |e_{s+1}\rangle \langle e_{s}|+ |e_s\rangle \langle e_{s+1}|\right). \nonumber\\
  \label{matIJ}
\end{eqnarray}
where  the vectors $|e_0\rangle,\ldots, |e_N\rangle$  are defined   in Eq. (\ref{vect}). 
The    stationary  Schr\"odinger equation in $\cH_N$,  Eq. (\ref{4Eq2}), is the   eigenvalue equation for the Jacobi matrix $\mathbf{J}_{N+1}$, where the energy eigenvalues are the   roots of the corresponding   characteristic equation, i.e., of the    Jacobi determinant
  \be
  \mathrm{det}\left(  \lambda\mathbf{I}_{N+1} - \mathbf{J}_{N+1}\right) =0.
  \en{EigProbJ}
It turns out that  the rescaled amplitudes are the principal minors of the  above  Jacobi determinant:
\be
 Y_{n} = \mathrm{det}\left(\lambda \mathbf{I}_n - \mathbf{J}_n\right).
 \en{YmatJ}
  Indeed, one can verify (by expanding the   matrix  determinant in Eq. (\ref{YmatJ}) over the last row/column) that   the sequence of the  determinants in  Eq. (\ref{YmatJ}) satisfies the recursion of Eq. (\ref{RecY}), where also  $Y_{0}=1$ (the null matrix) and $Y_1=\lambda$.   
Therefore, from Eqs. (\ref{psinYn-1}) and (\ref{YmatJ}) we obtain:
 \be
 \psi_n = \psi_0  \left(\prod_{s=0}^{n-1}\beta_s\right)^{-\frac12}\mathrm{det}\left(\lambda \mathbf{I}_n - \mathbf{J}_n\right). 
 \en{psidetJ}

 \subsection{The  stationary state }
 \label{sec4D}
 
 In the odd-dimensional invariant subspaces $\cH_N$, i.e, for even $N$, there is the zero eigenvalue $\lambda=0$ of the characteristic polynomial Eq. (\ref{YN}).  The corresponding rescaled quantum amplitude $Y_n(0)$ can be obtained from Eqs. (\ref{SolRecY})   and   (\ref{Gln_In}):
\begin{eqnarray}
 \label{Yn0}
&& Y_{2p+1}(0) = 0,\\
&& Y_{2p}(0)= (-1)^p\mathcal{G}^{(p)}_{2p-1}=(-1)^p\prod_{s=0}^{p-1}\beta_{2s}. \nonumber
\end{eqnarray}
The corresponding   $\psi_n$ of the   stationary state in $\cH_N$ with $N=2M$  can be found from  Eq.~(\ref{psinYn-1}).  The (non-normalized) stationary state with 
$\psi_0=1$    reads:
\begin{eqnarray}
\label{psi_n0}
&& \psi_{2p+1} = 0, \quad 0\le p\le M-1,\\
&& \psi_{2p} = (-1)^p \left(\frac{\prod_{s=0}^{p-1}\beta_{2s}}{\prod_{s=0}^{p-1}\beta_{2s+1}} \right)^\frac12, \quad 0\le p\le M. \nonumber
\end{eqnarray}

For example, consider the simplest two-mode  bosonic   model of Eq. (\ref{models}) with $m=S=1$ and arbitrary $k$,  with the interaction Hamiltonian $\hat{H} = \hat{a}^\dag \hat{b}^k + \hat{a}(\hat{b}^\dag)^k$.  In this case from  Eq. (\ref{beta_k}) we obtain
\be
\beta_n = (N-n)\prod_{j=1}^k(kn+j).
\en{beta_1}
The products in Eq. (\ref{psi_n0}) can be recast  in this case   as follows:
\begin{eqnarray}
&& \prod_{s=0}^{p-1}\beta_{2s} = 2^p(M)_p\prod_{s=1}^p\frac{([2s-1]k)!}{([2s-2]k)!},\nonumber\\
&& \prod_{s=0}^{p-1}\beta_{2s+1} = 2^p\left(M-\frac12\right)_p\prod_{s=1}^p\frac{(2sk)!}{([2s-1]k)!},
\label{betas}
\end{eqnarray}
 where we have used the notation   {$(n)_p = n(n-1)\ldots (n-p+1)$}  for the the falling factorial.  Thus the non-zero  amplitudes of the (un-normalized) stationary state read:
 \be
 \psi_{2p} = (-1)^p\left[\frac{(M)_p}{(M-1/2)_p} \prod_{s=1}^p \frac{\{([2s-1]k)!\}^2}{([2s-2]k)!(2sk)!}\right]^\frac12. 
 \en{psi_n0k} 
Observe that the first  factor  $ \frac{(M)_p}{(M-1/2)_p} $ in the square root in  Eq. (\ref{psi_n0k}) grows with $p$, whereas the second  factor decreases with $p$.  Hence, the    absolute values (un-normalized probabilities) $\psi_{2p}^2$ have local maxima   at $p=0$ (i.e., $n=0$) and  $p = M$ ($n= N$),  as illustrated in Fig. \ref{F1} for $k=2$.  
 \begin{figure}[h]
\begin{center}
      \includegraphics[width=.45\textwidth]{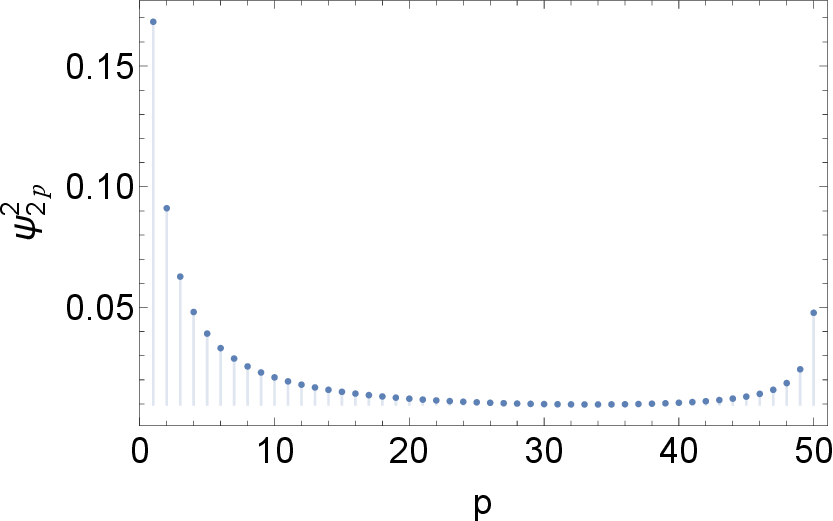} 
     \caption{ The non-zero squared amplitudes $\psi_{2p}^2$ of  the stationary state in Eq. (\ref{psi_n0k}) vs $p$ for $k=2$ and $N=100$ (the value $\psi_0^2=1$ was removed for better visibility of the distribution for $p>0$).    \label{F1} }
   \end{center}
\end{figure}

Quite similar distribution as in Fig. \ref{F1} one obtains for the multi-mode nonlinear boson models. For example, consider the model of Eq. (\ref{models}) with $m=k_s=1$ and $S=2$, i.e., the three-mode bosonic model with the interaction Hamiltonian $\hat{H} = \hat{a}^\dag \hat{b}_1\hat{b}_2 + \hat{a}\hat{b}^\dag_1\hat{b}^\dag_2$ and 
\be
\beta_n = (N-n)(n+1)^2. 
\en{bet3}
In this case  from Eq. (\ref{psi_n0}) one obtains the following  (un-normalized) stationary state in the invariant subspace $\cH_{2M}$    
\be
 \psi_{2p} = (-1)^p\left[\frac{(M)_p}{(M-1/2)_p}\right]^\frac12   \frac{(2p-1)!! }{(2p)!!},
 \en{psi_n03} 
 where, the first factor  (the same as in Eq. (\ref{psi_n0k})) grows with $p$, while the second one decreases with $p$. As the result,  we have  qualitatively the same behavior of the amplitudes as in Fig. \ref{F1}.

  \section{Discussion  }
\label{sec5}  
 
Exactly solvable models are considered highly useful in applications, provided that the methods leading to the exact solutions are sufficiently simple and the resulting expressions are amenable to further analysis. The usual approach to exactly solvable models, however, relies on rather sophisticated mathematical machinery, such as nonlinear deformations of group algebras \cite{GroupMeth,LieAlg3Boson,GroupMeth2,LieAlgGen1,LieAlgGen2}  and   the Quantum Inverse Scattering Method. \cite{QISM}.  
 
 Exactly solvable boson models are of particular interest not only because of the beautiful mathematical structures they embody but also due to their abundance in quantum physics, e.g., in quantum optics. Perhaps owing to the complicated mathematics involved, the physics literature—quite in parallel with analytical studies in mathematical physics—has primarily relied on numerical simulations and various approximations with poorly defined domains of applicability. Numerical simulations are demanding and resource-intensive, and only quite recently \cite{BeyondPA1,BeyondPA2} have numerical studies been able to  explore the strong-interaction or long-interaction-time regimes now achievable in experiments (e.g., Ref.~\cite{PumpDep}).
   
   On the other hand, perturbative series expansions involve powers of boson creation and annihilation operators, i.e., operators with unbounded norms, when one adopts the so-called parametric approximation for the strong pump mode. Such perturbative series lack well-defined domains of convergence in the rigorous mathematical sense, which limits their reliability—one is left to verify their validity only by comparison with experimental results.   A prominent example of this  is the spontaneous down-conversion process in quadratic nonlinear media and its generalizations to higher-order nonlinearities \cite{GenSqueez}. The standard perturbative approach based on the parametric approximation has been shown to be deficient for higher-order processes, as it leads to a divergence of the system state at a finite interaction time if the pump field is treated not as a separate quantum degree of freedom but merely as a parameter \cite{Buzek93,Drobny94}.  It is, in fact, by chance that no divergence occurs in the quadratic case, where the standard parametric approximation yields the squeezed state of light \cite{ReviewSq}, which remains regular in norm. Nevertheless, the domain of validity of the standard parametric approximation had long remained undefined. Only very recently has this issue been addressed through direct comparison with a non-perturbative asymptotic  approach \cite{ourPRA}. The latter was constructed upon the existence of an exact solution previously derived for the evolution of a special initial state in Ref.~\cite{My1}.

   It was therefore deemed necessary to generalize the exact approach of Ref.~\cite{My1} to its most general form—the  task accomplished in the present work. The  previous  approach has been extended in this work to the arbitrary state evolution and, additionally,  to  analysis of the energy spectrum, while  relying only on elementary algebra, continued fractions, and the    recurrence relations. Remarkably, this framework applies to a broad class of exactly solvable boson models sharing two defining features, as discussed in Section~\ref{sec2}.

The limitations of the current approach stem from its very generality: since the solution is fully characterized by a single polynomial parameter, $\beta_n$ in Eq.(\ref{Psi_n}) of Section~\ref{sec2}, distinct physical regimes correspond to different choices of this polynomial (see Ref.~\cite{My1} for details), even though the form of the solution remains the same. Thus, to advance the understanding of specific models, one must go beyond generality and focus on a particular  $\beta_n$ representing a distinct physical situation. Clearly, this task requires development of  additional approaches, supplementing the present general framework. The significance of the present work lies in providing a rigorous, divergence-free framework for such future analyses of particular models.

  \section{Conclusion}
\label{sec6}  
  
The analytical solutions to both the state-evolution and energy-spectrum problems have been obtained for a broad class of bosonic models sharing two key features:
(i) the interaction Hamiltonian is a sum of two Hermitian-conjugate ladder operators, and
(ii) the Hilbert space decomposes into finite-dimensional invariant subspaces under this interaction.
Consequently, a single polynomial function of the discrete index labeling the basis states within each invariant subspace fully characterizes the model, governing both its dynamical evolution and spectral properties.

The   explicit solution to the arbitrary state evolution problem has been found  in the form of an infinite time-series expansion whose coefficients are nested sums involving the characterizing  polynomial. The characteristic polynomial for the energy eigenvalues has coefficients of a similar structure.
The eigenstate amplitudes have been represented equivalently as continued fractions, as products of M\"obius-group matrices in the projective space, and as the principal minors of the associated Jacobi matrix.
The  results  have important area of applications to  the  nonlinear quantum-optical models,  including  the $k$-photon down-conversion, which play a central role in modern quantum technologies.

An important open problem is  the  asymptotic limit   as the dimension of the invariant subspace (for instance, the mean photon number in the pump mode) tends to infinity.  This problem has important  immediate application to the spontaneous down-conversion process, the workhorse of modern quantum optics. In the asymptotic  limit  one can expect to obtain a simple  analytical approximation  which would reduce to the      common parametric approximation  under the appropriate conditions \cite{ourPRA}.  The present work  provides a solid foundation for addressing this question in the future.

\section{acknowledgements}

The author acknowledges financial support, grant number 307507/2023-8,   from  the National Council for Scientific and Technological Development (Conselho Nacional de Desenvolvimento Cient\'ifico e Tecnol\'ogico) of Brazil.  

\newpage
\appendix
 \section{Proof of Eq. (\ref{Amk_EX}) of section \ref{sec2}}
\label{appA}
We will use the  following    expansion (a generalization of the   binomial theorem)
\be
(\hat{A} + \hat{A}^\dag)^m = \sum_{l=0}^m \sum_{P_\mu}P_\mu\{\hat{A}^l (\hat{A}^\dag)^{m-l}\},
\en{Ae1}
where $P_\mu$ is the permutation operator acting on the product of $m$ factors, where $\mu$  runs over   different orderings  of $m$ factors, with $l$ factors of one type and $m-l$ of the other  type (there are  $\binom{m}{l}$ of different orderings). Using Eq. (\ref{AAd}) and (\ref{FA}) we get from Eq. (\ref{Ae1}):
\begin{eqnarray}
\label{Ae2}
&&(\hat{A} + \hat{A}^\dag)^m (\hat{A}^\dag)^k = \sum_{l=0}^{\left[\frac{m}{2}\right]}(\hat{A}^\dag)^{k+m-2l} a^{(l)}_{m,k}(\hat{n})\nonumber\\
&& + \sum_{l=\left[\frac{m}{2}\right]+1}^{m}  (\hat{A}^\dag)^k\hat{A}^{ 2l-m}  a^{(l)}_{m,k}(\hat{n}),
\end{eqnarray}
where the scalar functions $a^{(l)}_{m,k}(n)$ are   combinations of $\beta_n$ of Eq. (\ref{Psi_n}).  Using that  $\hat{A}^p|\Psi_0\rangle=0$ for $p>0$  and 
introducing the scalars $a^{(l)}_{m,k} \equiv a^{(l)}_{m,k}(0)$, we obtain:
\begin{eqnarray}
\label{Ae3}
&&(\hat{A} + \hat{A}^\dag)^m (\hat{A}^\dag)^k |\Psi_0\rangle = \sum_{l=0}^{\left[\frac{m}{2}\right]}(\hat{A}^\dag)^{k+m-2l} a^{(l)}_{m,k} |\Psi_0\rangle \nonumber\\
&& + \sum_{l=\left[\frac{m}{2}\right]+1}^{\left[\frac{k+m}{2}\right]}  (\hat{A}^\dag)^{k+m- 2l}  a^{(l)}_{m,k}|\Psi_0\rangle,
\end{eqnarray}
where   the second sum  is non-zero for $2l\le m+k$. Combining the two sums in Eq.~(\ref{Ae3}) and using that 
$a^{(l)}_{m,k}(n)$ is nonzero only for  $l\le m$,  one arrives  arrive at Eq. (\ref{Amk_EX}) of section \ref{sec2}. 



\begin{thebibliography}{99}


\bibitem{GroupMeth} V. P. Karassiov and A. B.Klimov, An algebraic approach for solving evolution problems in some nonlinear quantum models, 
Phys. Lett.  A \textbf{191}, 117  (1994). 

\bibitem{GroupMeth2}  V. P. Karassiov,  Symmetry approach to reveal hidden coherent structures in Quantum Optics. General outlook and examples, 
J. Russian  Laser Research, \textbf{21},  370 (2000).
 
\bibitem{LieAlg3Boson} V. P. Karassiov,  A. A. Gusev, and  S. I. Vinitsky,  Polynomial Lie algebra methods in solving the second-harmonic generation model: some exact and approximate calculations, Phys.  Lett.  A  \textbf{295},  247 (2002). 

\bibitem{LieAlgGen1}   Y.-H. Lee, W.-Li Yang, and Y.-Zh.  Zhang, Polynomial algebras and exact solutions of general quantum nonlinear optical models I: two-mode boson systems, J. Phys. A: Math. Theor. \textbf{43},   185204 (2010). 

\bibitem{LieAlgGen2}   Y.-H. Lee, W.-Li Yang, and Y.-Zh.  Zhang,   Polynomial algebras and exact solutions of general quantum nonlinear optical models: II. Multi-mode boson systems,  J. Phys. A: Math. Theor. \textbf{43},   375211 (2010). 




\bibitem{SpinSystems} V. V. Ulyanov and  O. B. Zaslavskii, New methods in the theory of quantum spin systems, Phys. Rep. \textbf{216}, 179 (1992). 

\bibitem{3BosonBethe} V. A. Andreev and O. A. Ivanova, The dynamics of three-boson interaction and algebraic Bethe ansatz, Phys. Lett. A. \textbf{171}, 145 (1992). 

\bibitem{QISM} V. E. Korepin,   N. M. Bogoliubov, and A. G. Izergin, \textit{Quantum Inverse Scattering Method and Correlation Functions} (Cambridge University Press, 1993). 



\bibitem{NonlOpt} R. W. Boyd, \textit{Nonlinear Optics} (Academic Press, Elsevier 2008). 



\bibitem{GenSqueez}  S. L. Braunstein and R. I. McLachlan, Generalized squeezing,  Phys. Rev. A \textbf{35}, 1659 (1987). 

\bibitem{3phDC}   W. S. Chang, C. Sab\'in, P. Forn-D\'iaz,  F. Quijandr\'ia,  A. M. Vadiraj, I. Nsanzineza, G. Johansson, and C. M. Wilson, Observation of Three-Photon Spontaneous Parametric Down-Conversion in a Superconducting Parametric Cavity,  Phys. Rev. X  \textbf{10}, 011011 (2020). 

\bibitem{ReviewSq} R. Loudon and P. L. Knight, Squeezed light, J. Mod. Opt. \textbf{34}, 709 (1987). 

\bibitem{Couteau18} C. Couteau,  Spontaneous parametric down-conversion, Contemporary Physics, \textbf{59},  291 (2018). 



 \bibitem{Zhang21} C. Zhang, Y.-F. Huang, B.-H. Liu, C.-F. Li,  and G.-C. Guo, Spontaneous Parametric Down-Conversion Sources for Multiphoton Experiments, 
Adv. Quantum Technol.  \textbf{4},  2000132 (2021). 

\bibitem{ClasskSq} R. A. Fisher,     M. M. Nieto, and V. D. Sandberg, Impossibility of naively generalizing squeezed coherent states, Phys. Rev. A \textbf{29}, 1107 (1984). 

\bibitem{WKB4GenFun} G. Scharf,  Time Evolution of a Quantum Mechanical Maser Model,  Annals  of Physics, \textbf{83}, 71 (1974). 


\bibitem{Hillery84}  M. Hillery and M.S. Zubary, Path-integral approach to the quantum theory of the degenerate parametric amplifier, Phys. Rev. A \textbf{29}, 1275 (1984). 

\bibitem{WKBpump} G. Scharf and D. F. Walls, Effect of pump quantization on squeezing in parametric amplifier,  Opt. Comm. \textbf{50}, 245 (1984). 

\bibitem{Crouch88} D. D. Crouch and S. L. Braunstein, Limitations to squeezing in a parametric amplifier due to pump quantum fluctuations, Phys. Rev. A \textbf{38}, 4696 (1988). 

\bibitem{Reid88} M. D. Reid and P. D. Drummond, Quantum Correlations of Phase in Nondegenerate Parametric Oscillation, Phys. Rev. Lett. \textbf{60}, 2731 (1988). 

\bibitem{Drobny92} G. Drobn\'y and I. Jex, Quantum properties of field modes in trilinear optical processes, Phys. Rev. A \textbf{46}, 499 (1992). 

\bibitem{Buzek93}  V. Buzek and G. Drobn\'y, Signal-pump entanglement in quantum  $k$-photon down-conversion, Phys. Rev. A \textbf{47}, 1237 (1993). 

\bibitem{Drobny94}  G.  Drobn\'y and V. Buzek, Fundamental limit on energy transfer in $k$-photon down-conversion,  Phys. Rev. A \textbf{50}, 3492 (1994). 

\bibitem{Hillery95}  M. Hillery, D. Yu, and J. Bergou, Effect of the pump state on the dynamics of the parametric amplifier, Phys. Rev. A \textbf{52}, 3209 (1995)


\bibitem{BeyondPA1} W.  Xing and T. C. Ralph, Pump depletion in optical parametric amplification, Phys. Rev. A \textbf{107}, 023712 (2023). 

\bibitem{BeyondPA2} K.  Chinni and N. Quesada, Beyond the parametric approximation: Pump depletion, entanglement, and squeezing in macroscopic down-conversion, 
Phys. Rev. A  \textbf{110}, 013712  (2024). 


 \bibitem{My1} V. S. Shchesnovich, Exact solution for a class of quantum models of interacting bosons,   J. Phys. A: Math. Theor. \textbf{58}, 115204 (2025). 

\bibitem{Hessen}  C. P. Huang, Computing Powers of Arbitrary Hessenberg Matrices, Linear Algebra  Appl.  \textbf{21}, 123  (1978). 


\bibitem{BookCF}  S. Khruschev, \textit{Orthogonal Polynomials and Continued Fractions. From Euler’s Point of View} (Cambridge University Press, 2008). 


\bibitem{ourPRA}    D. B. Horoshko  and V. S. Shchesnovich, Isoenergetic model for optical down-conversion and error-specific limits of the parametric approximation,   Phys. Rev. A \textbf{112}, 033706 (2025). 
 
 
 \bibitem{PumpDep} J. Fl\'orez, J. S. Lundeen, and  M. V. Chekhova, Pump depletion in parametric down-conversion with low pump energies, Optics Letters \textbf{45},   4264 (2020). 

\end{thebibliography}
\end{document}